\documentclass[aps,pre,superscriptaddress,twocolumn]{revtex4} 
\usepackage{mathtools}
\usepackage{lipsum}
\usepackage{graphicx}
\usepackage{dcolumn}
\usepackage{bm}
\usepackage{xcolor}
\definecolor{sapphire}{RGB}{15,82,186}
\usepackage[colorlinks=true,citecolor=sapphire,linkcolor=sapphire,urlcolor=sapphire]{hyperref}

\begin{document}

\title{{Self-avoiding tethered surfaces are always flat}} 

\author{A. D. Chen}
\affiliation{Department of Chemistry, Columbia University, 3000 Broadway, New York, NY 10027}
\author{M. C. Gandikota}
\affiliation {International Centre for Theoretical Sciences, Tata Institute of Fundamental Research, Bengaluru 560089, India}
\author{M. J. Kim}
\affiliation{Department of Chemistry, Columbia University, 3000 Broadway, New York, NY 10027}
\author{A. Cacciuto}
\email{ac2822@columbia.edu}
\affiliation{Department of Chemistry, Columbia University, 3000 Broadway, New York, NY 10027}

\begin{abstract}
\noindent
The scaling behavior of fully flexible elastic tethered surfaces has been debated for decades. Some theories predict that self-avoiding surfaces would crumple in the absence of bending rigidity, while most simulations suggested that they would remain flat. 
Recent simulations on ideal membranes with lattice perforations suggest that systematically removing surface area from a membrane may provide an alternative way to crumpling self-avoiding surfaces. We perform extensive numerical simulations of two models of fully flexible elastic tethered surfaces in which self-avoidance can be systematically and continuously tuned to the ideal limit. We show that in the thermodynamic limit, these surfaces remain flat with a size exponent $\nu=1$ for any finite degree of self-avoidance, with or without membrane perforations.
 
\end{abstract}
\maketitle
\section{Introduction} A longstanding problem in statistical mechanics concerns the shape of elastic tethered surfaces, fundamental structural elements of a myriad of diverse biologic and synthetic materials.
Examples include connective or epithelial tissues~\cite{tissue}, red blood cells~\cite{Schmidt1993Feb,Lux2016Jan},  viruses~\cite{Lidmar2003Nov}, pollen~\cite{Pollen}, as well as 
graphene sheets~\cite{Stankovich2006Jul},  polymer~\cite{Huang2007Aug} and nanoscale DNA films~\cite{dnafilm}. 
In contrast to fluid surfaces, the fixed connectivity of tethered membranes imposes an energetic penalty for shear deformations. 
The phase behaviour of {\it ideal} tethered surfaces is well understood. At low bending rigidities, the surface is found in a crumpled state that is characterized by a radius of gyration $R_g$ that scales as $R_g^2\sim \log (N)$ where $N$ is proportional to the area of the surface. 
For sufficiently large bending rigidities, the surface remains flat with $R_g^2\sim N$. The critical bending rigidity is found to be merely $0.33\, k_{\rm B}T$~\cite{kantor_crumpling_1987}. The subject is reviewed in~\cite{NelsonBook}.

This question remains unsettled for fully flexible self-avoiding tethered surfaces. 
The generalization of Flory theory~\cite{Gennes1979Nov} (greatly successful for understanding the size scaling of self-avoiding polymers to surfaces) predicts $R_g^2\sim N^\nu$, with a size exponent $\nu=(D+2)/(d+2)$. Here, $D$ is the intrinsic dimension of the object ($D=1$ for a polymer and $D=2$ for a surface), and $d$ is the dimension of the embedding space. For physical tethered surfaces ($d=3$), this implies that $R_g^2\sim N^{\frac{4}{5}}$~\cite{domb2001,paczuski1988}. Most early numerical simulations of tethered self-avoiding elastic sheets have, however, indicated that these surfaces remain overall extended (flat) with a size exponent along their longitudinal directions equal to $\nu\approx 1$, even in the absence of an explicit bending rigidity~\cite{plischke1988,abraham1991folding,bowick2001,abraham1989,Boal1989Sep}.
The existence of a self-avoiding crumpled phase characterized by a size exponent $\nu=4/5$ for large tethered membranes in equilibrium and in the absence of bending energy,  has remained a point of contention, even in light of  sophisticated two-loop renormalization group calculations~\cite{DavidRg,wiese2000} which pointed to an asymptotic limit compatible with the Flory exponent.
Early experiments on graphite-oxide sheets suspended in poor solvents~\cite{Wen1992Jan} appeared to indicate the presence of a crumpled phase as the solvent quality improved. However, conducting experiments under these conditions has proven challenging, and later studies have suggested results to the contrary~\cite{Koltonow2017-ro}.
Indeed, as discussed above, just a fraction of a $k_{\rm B} T$ in bending energy would be sufficient to flatten an ideal membrane. At the same time, a small amount of attractive dispersion forces can drive the surface into a folded compact state. Interestingly, thin elastic spherical shells in the presence of explicit attractive forces were initially reported to have a Flory exponent compatible with a crumpled phase for a narrow range of temperatures~\cite{liu1992}, but simulations with larger shells did not find such an intermediate regime~\cite{grest1994}. Overall, much of the numerical work conducted on this subject has been limited to relatively small system sizes, making the results susceptible to significant finite-size effects.

One way to reliably obtain a crumpled phase out of elastic thin surfaces is by quickly compressing them using a large external force~\cite{kantor1988,gomes1987,kantor1987,gomes1989} or by rapidly dehydrating graphene-oxide nanopaper~\cite{ma2012}. The destabilization of these flat phases \textit{via} compression has also been studied numerically~\cite{vliegenthart2006forced}. More recently, it has been shown that active forces are capable of crumpling an elastic shell with a size exponent compatible with the expected Flory exponent of $4/5$~\cite{gandikota2}, but active forces were unable to crumple an elastic sheet~\cite{Gandikota1}.
The collapse of a self-avoiding membrane has been shown to occur upon performing a sufficient number of systematic parallel cuts across the whole length of an elastic sheet. However, the surface only collapses when the number of cuts becomes so large that the membrane effectively reduces to a thin square frame linked by parallel self-avoiding polymer strips~\cite{Gandikota3}. In this regime, with virtually no lateral connectivity between the strips, the system can hardly be described as a self-avoiding surface.

Attempts have been made to decrease the strength of self-avoidance by decreasing the diameter of the self-avoiding particles that endow the membranes their self-avoidance ~\cite{abraham1989,Boal1989Sep}. It has been argued that for large enough system sizes, the flat phase persists even for the smallest particle diameters. An even stronger argument was made by removing the particles all together and imposing self-avoidance by forbidding any degree of overlap between the triangles (plaquettes) defining the surface \textit{itself}~\cite{kroll1993floppy,bowick2001}. These simulations also pointed to the robustness of the flat phase in self-avoiding membranes.

\begin{figure}[t]
    \centering
    \includegraphics[width = 0.35\textwidth]{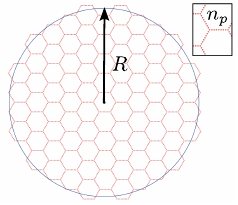}
    \caption{Sketch of the surface model with polymers of length $n_p=8$ containing 85 hexagonal cells. The inset shows the detail of the connections in each node of the membrane. The membrane conformation is defined  by two parameters: $n_p$, the number of monomer units per polymer segment between adjacent nodes, and $R$, the radius of a circle centered on the surface, which determines the extent of the network by excluding any hexagons whose centers lie outside this circular boundary. }
    \label{fig1A}
\end{figure}
Yet, this is not the only way to tunably weaken self-avoidance. Recent studies have demonstrated that introducing periodic perforations in ideal, {\it phantom}, tethered membranes reduces the critical temperature associated with the crumpling transition~\cite{yllanes2017thermal}, and suggested that 
distant self-avoidance in the presence of a lattice of perforations may 
be the way to crumple these surfaces; they argued that the crumpled ring polymer phase is known to survive the presence of self-avoidance~\cite{Gennes1979Nov}.
In this paper, we explore this scenario, and using extensive numerical simulations, where---to maximize the area removed---we model our surface as a polymer network with hexagonal symmetry, constructed by  connecting fully flexible polymer chains of uniform length at each node of the surface. We show that even in this case the self-avoiding surface remains flat, independent of the amount of area removed.

\section{Methods}
We model the membrane as a fishnet network where each node particle serves as a three-fold coordinated vertex
linking three distinct polymer strands containing each $n_p$ monomers, generating a two-dimensional network with hexagonal symmetry. A surface characterized by $n_p=8$ consisting of 85 hexagonal cells is depicted in Fig.~\ref{fig1A}.

\begin{figure}[t]
    \centering
    \includegraphics[width = 0.45\textwidth]{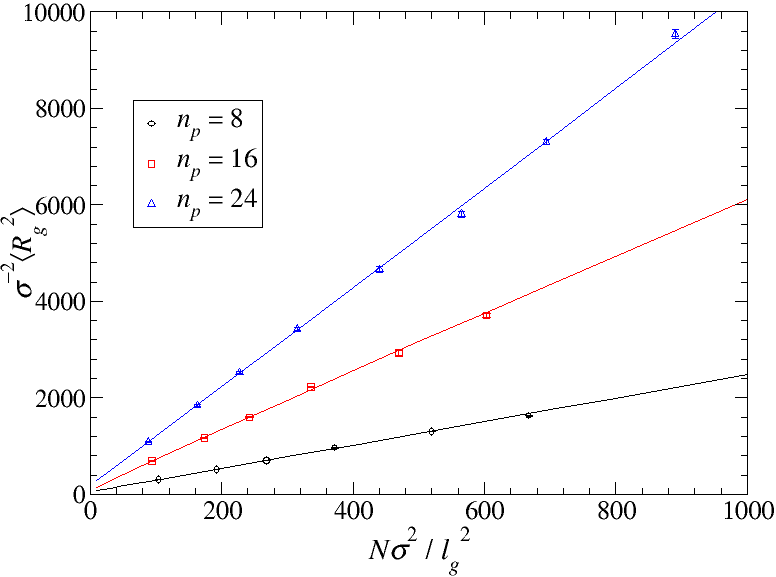}
    \caption{Scaling of the square of the radius of gyration, $R_g$, as a function of total number of nodes rescaled by the characteristic size of a self-avoiding polymer, $l_g=\sigma\, n_p^{0.588}$, for fish-net  systems characterized by $n_p=8,16$ and $24$ monomers per polymer link. The solid lines are the fits to the data with the functional form $f(x)=a\,x^c+b$. The powers are $c=0.98\pm0.02$, $0.97\pm0.05$, and $1.01\pm0.03$ for $n_p=8, 16$ and 24 respectively.}
    \vspace*{-0.5cm}
    \label{fig2}
\end{figure}
The polymer monomers of diameter $\sigma$ are held together by the harmonic potential $U_s=K_s\,(|\Delta r|-d)^2$ where $|\Delta r|$ is the distance between any two connected monomers, $d=1.6\,\sigma$ is the equilibrium distance, and $K_s=160\,k_{\rm B}T/\sigma^2$ is the stretching constant. 
Excluded volume between any two of the $N$ monomers making up the surface is enforced \textit{via} a WCA potential~\cite{weeks_role_1971} 
with the energy scale $\varepsilon=k_{\rm B}T$. 
The statistical properties of the system are obtained using Langevin dynamics simulations
\begin{equation}
    m\frac{d\textbf{v}_i}{dt} = \textbf{f}_i - \gamma\,\textbf{v}_i + \sqrt{2D\gamma^2}\, \bm{\xi}_i(t)
\end{equation}
where $m$ is the mass of any given node particle, $i$ represents the index of a given particle. The velocity of each particle is $\textbf{v}_i$, $\gamma$ is the translational friction coefficient and $D = k_{\text{B}}T\gamma^{-1}$ is the translational diffusion constant. 
$\textbf{f}_i = -\partial U/\partial \textbf{r}_i$ are the forces among the particles. All simulations have been carried out using the numerical packages HOOMD~\cite{Glaser2015-mc} and LAMMPS~\cite{ plimpton_fast_1995}. The units of length, time, and energy are set to be $\sigma$, $\tau=\sigma^2D^{-1}$, and $k_{\text{B}}T=\beta^{-1}$, respectively. The time step is set to $\Delta t = 0.02\,\tau$, and $\gamma=1$.
All simulations are run for a minimum of $2\times 10^9$ steps and a maximum of $8\times 10^9$ steps.

\section{Results}
\subsection{Fishnet Networks} 
\begin{figure}[h]
    \centering
    \includegraphics[width = 0.45\textwidth]{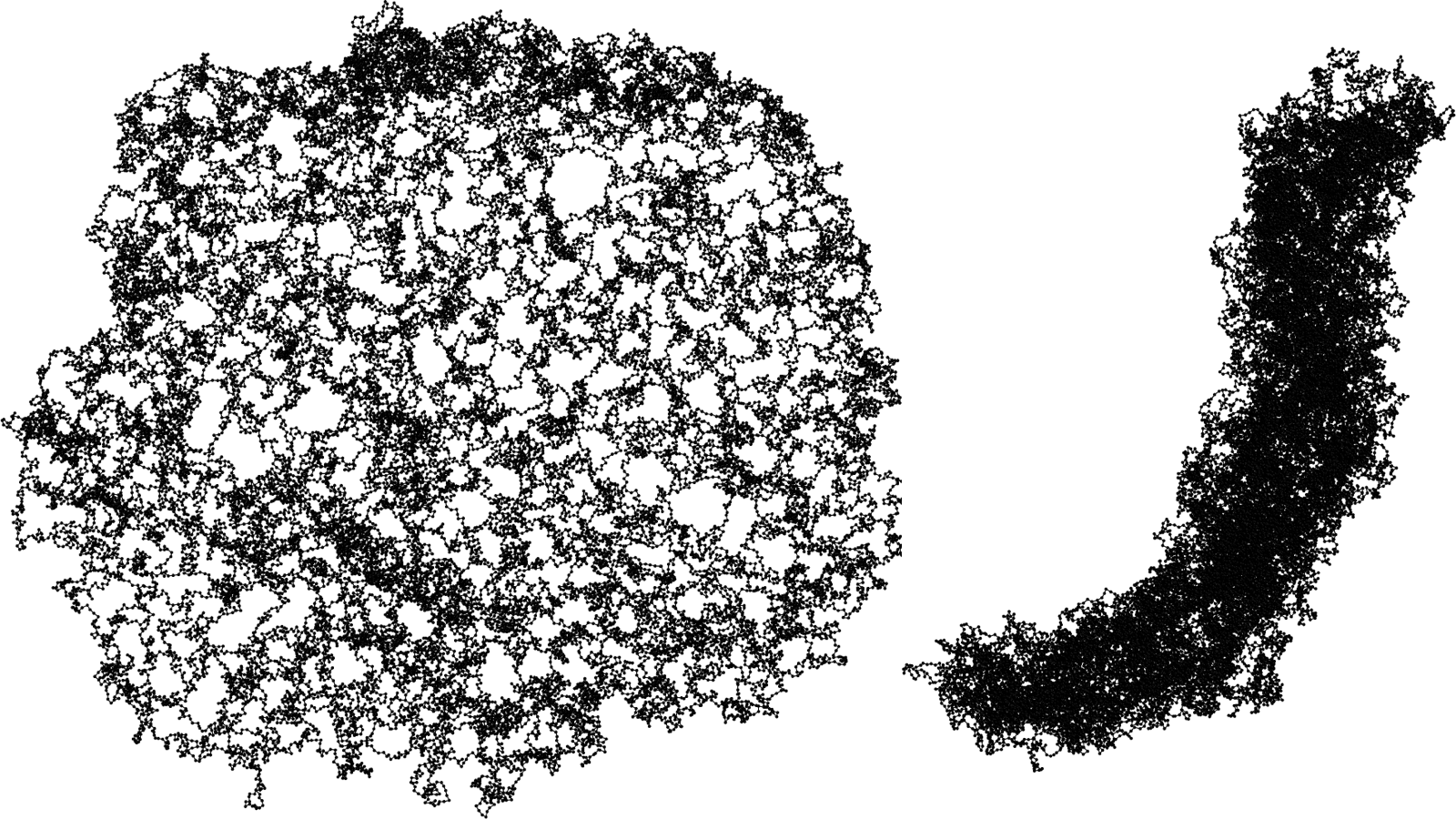}
    \caption{Top and side view of a thermalized configuration of the largest surface simulated in our study. Each link contains $n_p=24$ particles for a total of $N=37374$ particles.}
    \label{figB}
\end{figure}

We begin our analysis by measuring the scaling of the radius of gyration of the membrane
as a function of the total number of particles $N$ for different values of $n_p=8, 16$ and 24.
Remarkably, even for our sparsest surfaces, $n_p=24$, we observe a power-law scaling of the squared radius of gyration with the system size, consistent with flat membrane scaling. 
Specifically, we fit $R_g^2$ as function of $N\sigma^2$ rescaled by 
$l_g^2$, where $l_g=\sigma n_p^{\nu_p}$ with $\nu_p=0.588$~\cite{rubinstein2003polymer}, is an estimate of the radius of gyration of each polymer.
Essentially, $N\sigma^2/l_g^2$ is an estimate of the number of polymer blobs in the system.
Figure~\ref{fig2} is one of the main results of this paper, and suggests that distant self-avoidance is sufficient to stabilize the flat phase of the surface. Representative snapshots of a thermalized configuration is shown in Fig.~\ref{figB}.

One could, in principle, keep on increasing $n_p$ and further dilute the surface. However, even on the latest GPUs, it would be very challenging to extract statistically independent configuration from simulations with  larger values of $n_p$, requiring more than $10^5$ monomers, and far exceeding $10^{10}$ iterations of the molecular dynamics integrator. 
It could be argued, as is for early numerical simulations of self-avoiding surfaces that dealt with relatively small number of surface nodes, that for larger values of $n_p$ the surface could acquire a size exponent smaller than 1, and closer to the value expected from the Flory arguments. 

To overcome this issue, we also considered a different model for our polymer network surface, one that captures what the effective behavior the surface would be if dealing with extremely long chains. This can be done by
recognizing  that, in this limit, the surface  can be thought of as a network built out of polymer blobs. It is crucial to recognize that the free energy cost to overlap two fully flexible polymer blobs is finite, and it was measured to be approximately $2\,k_{\rm B}T$, \textit{independent} of the length of the polymer~\cite{Grosberg82,Louis2000Sep} (a simple scaling argument is given in the Appendix~\ref{appendix}). This result suggests that, even in the $n_p\rightarrow\infty$ limit, a self-avoidance cost of $2\,k_{\rm B}T$ will remain in place between the units of the effective membrane with polymer blobs as surface nodes, setting a clear physical lower limit for the blobs' interaction energy. 
This mapping allows us to effectively capture the behavior of extremely sparse surfaces, $n_p\rightarrow\infty$.

\subsection{Soft Self-avoiding Tethered Membranes} 
\begin{figure}[h]
    \centering
    \includegraphics[width = 0.45\textwidth]{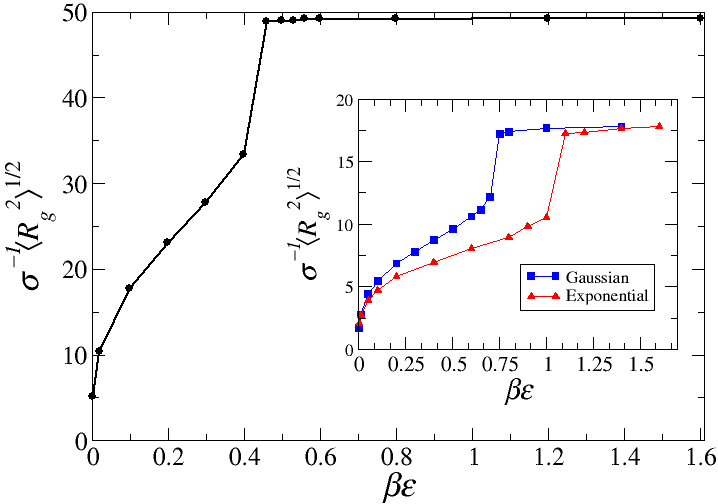}
    \caption{Radius of gyration, $R_g$, of the soft-sphere model as a function of the degree of self-avoidance $\beta\varepsilon$ for a $N=80\times80$ surface. The inset is the same plot for a $N=30\times30$ membrane the elements of which interact with a Gaussian and with an Exponential potential as discussed in the main text.}
    \vspace*{-0.5cm}
    \label{fig3}
\end{figure}

To re-frame our problem for large $n_p$ we now consider a membrane model built out of ultra soft particles, each corresponding to a polymer blob, and ask what is the lowest amount of softness below which the flat phase is destabilized. In this model, the bonds connecting the blobs of diameter $\sigma$ are identical to those in the previous model, the difference being that we now use a simple hexagonal geometry for the whole membrane (see Fig.~\ref{fig1A}), and the excluded volume interaction is enforced \textit{via} the potential, 
\begin{equation}\label{usa}
U_\text{SA}(r) =
\begin{cases}
\frac{\varepsilon}{2}\left [1+\text{cos}\left(\frac{\pi r}{\sigma}\right)\right], &r\leq\sigma \\
0, & r>\sigma.
\end{cases}
\end{equation}
This potential has the property of having a well defined blob diameter,  is smooth at $r=\sigma$, and has a {\it finite} energy cost, controlled by $\varepsilon$, for full particle overlap. The parameter $\varepsilon$ can therefore be used to continuously interpolate this model between an ideal and a self-avoiding tethered surface.
Figure~\ref{fig3} shows how $R_g$ depends on $\varepsilon$ for an $N=80\times80$ soft membrane. 

Here, we see that the radius of gyration of the surface experiences a sudden drop upon decreasing $\varepsilon$ sufficiently.
This result is very important because it implies that the destabilization of the flat phase occurs as soon as the overlapping  energy becomes smaller than $\varepsilon^*\approx 0.45\,k_{\rm B}T$.
Given the  $2\,k_{\rm B}T$ lower bound for the overlapping free energy between two polymers, our results suggest that the open cell surface discussed earlier will not be destabilized even in the limit for $n_p\rightarrow \infty$. A careful analysis of the membrane morphology for values of $0.1\, k_{\rm B}T\!<\!\varepsilon\!<\!0.45\,k_{\rm B}T$ reveals that surprisingly the overall shape of the surface remains flat even for $\varepsilon<\varepsilon^*$, and the sudden decay in $R_g$ when decreasing $\varepsilon$ is not the result of a crumpling transition, but is caused by the inner \textit{creasing} of the surface which, however, maintains an overall extended shape with $R_g^2\sim N\sigma^2$. We also tested this result with two alternative ultra-soft interaction potentials: (a) $U(r)=\varepsilon\,e^{-r/\sigma}$ and 
(b) $U(r)=\varepsilon\, e^{-r^2/\sigma^2}$, both cutoff at $r_c=\sigma$. Apart from the specific location of $\varepsilon^*$, which, however, is always below $2\,k_{\rm B}T$, the overall behavior is consistent with that observed for our potential (see inset of Fig.~\ref{fig3}).

\subsection{Creasing} The dynamics of how the surface reduces its area when starting from a perfectly flat conformation is characterized by the formation of small, non-aligned, creases within which particles strongly overlap with each other. The creases typically begin to form from the edges of the surface, but eventually also develop at its core and grow to span the surface, generating an intricate network. The net result is a surface that is smaller in lateral length, but is thicker because of the multi-particle occupancy per site.
The smaller the $\varepsilon$, the deeper and more numerous the creases are, and thus the thicker and smaller the resultant surface is (See S.I. for a movie of the surface dynamics).

\begin{figure}[t]
    \centering
    \includegraphics[width = 0.45\textwidth]{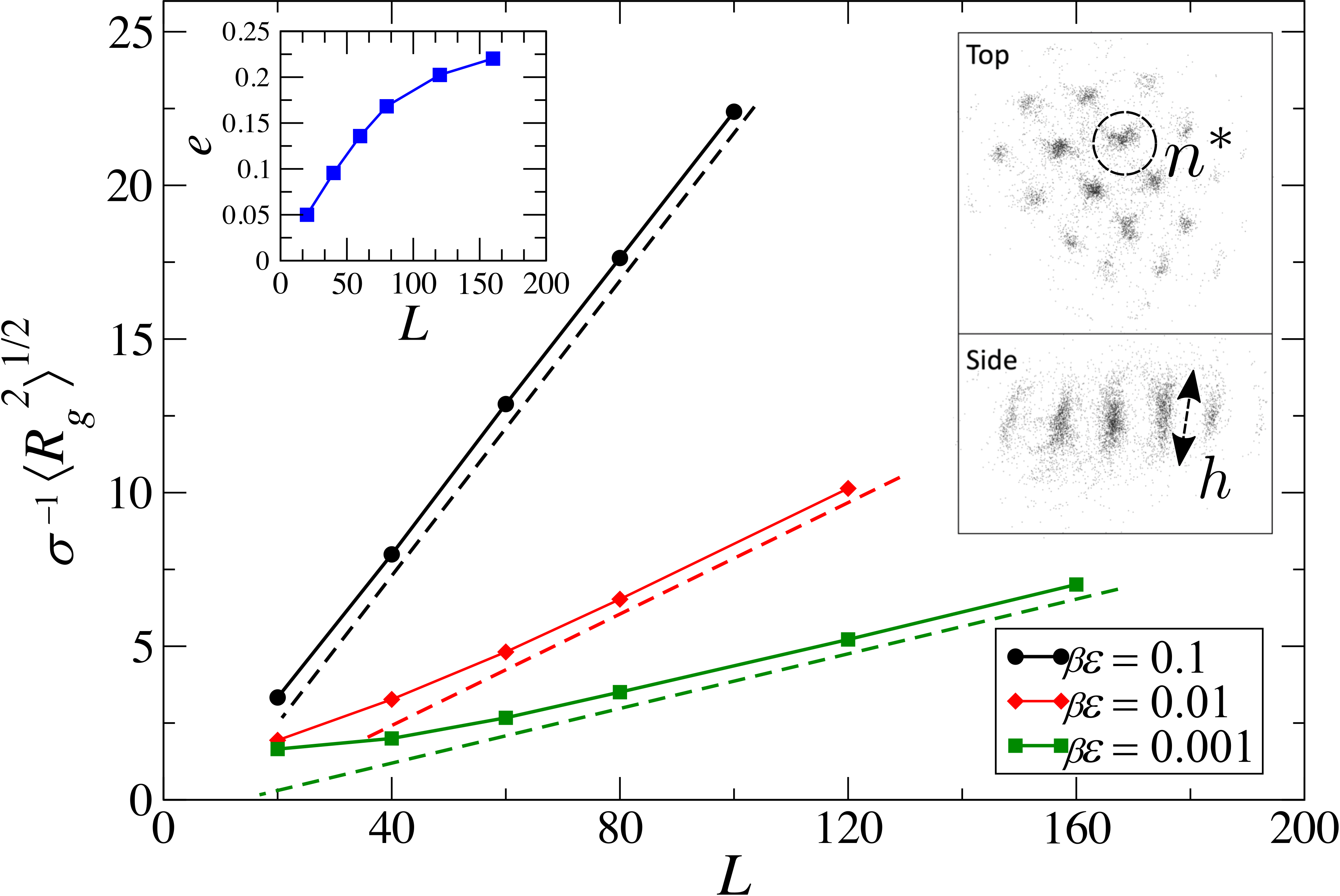}
    \caption{Radius of gyration of the soft-sphere model as a function surface length $L$ for various $\varepsilon$. The solid lines are guides to the eye. The straight dashed lines serve to highlight the onset of the linear dependence of the curves. The left inset  shows the asphericity $e$ of the surfaces as a function of $L$ for $\beta\varepsilon=0.001$. The right inset shows top and side views of a surface configuration for $\beta\varepsilon=10^{-5}$. Particles sizes have been greatly reduced to highlight their site multi-occupancy.}
    \vspace*{-0.5cm}
    \label{fig4}
\end{figure}

Once the system has attained equilibrium, it is important to explore the size dependence of $R_g$ in these creased membranes with $\beta\varepsilon\leq 0.1$. Fig.~\ref{fig4} shows the result of this analysis for $\beta\varepsilon=10^{-1}, 10^{-2}$ and $10^{-3}$. While in the first case, we recover a simple linear behavior $R_g\sim L$, the curves for smaller values of $\varepsilon$ suggest a more complex size dependence. 
Specifically, for sufficiently small values of $N$ the membrane is fairly isotropic, but it eventually expands laterally for large $N$. This trend is well characterized by the asphericity parameter $e$~\cite{rudnick_aspherity_1986}, which grows from $e\approx 0$ for small $N$ to $e\approx 1/4$ for large $N$ (see onset of Fig.~\ref{fig4}). The former value is compatible with an isotropic structure while the later is indicative of a planar structure. Crucially, the size exponent of the surface for large $N$ is again linear with $L$ - the expected value for a flat surface.
A careful look at the configurations for small $\beta\varepsilon$ reveals  a clear ordering on the surface, where  several particles cluster around the surface  nodes (multi-occupancy)~\footnote{Notice that none of the soft potentials used for self-avoiding interactions lead to the formation of cluster crystals in dense solutions of soft particles~\cite{Likos_cluster}} with a given positional spread along the perpendicular direction to the surface at that point. This is clearly shown in the right inset of Fig.~\ref{fig4}, which shows top and side view for a configuration at $\beta\varepsilon=10^{-5}$. The smaller $\varepsilon$ the larger the vertical spread and the smaller the overall size of the surface.

\begin{figure}[t]
    \centering
    \includegraphics[width = 0.45\textwidth]{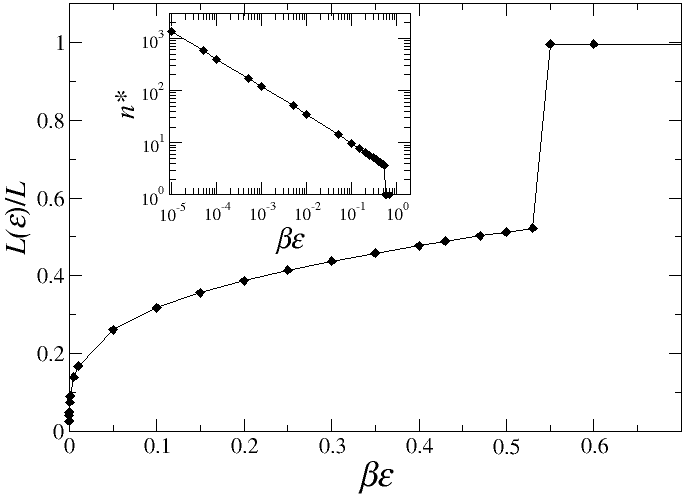}
    \caption{Normalized lateral size of the membrane $L(\varepsilon)/L$ as a function of the the strength of the self-avoidance $\beta\varepsilon$ as obtained from the theoretical multi-occupancy model. The inset shows the number of monomers $n^*$ overlapping their position on each node of the surface as a function of $\beta\varepsilon$ in a log-log scale.}
    \label{theory}
\end{figure}
We can then think of the folded surface as an effectively smaller surface where on average, $n$ particles sit on each of its nodes, and one can estimate the value of $n$ by imposing the stability condition that the repulsion energy between two fully overlapping sites, each containing $n$ particles, must be equal to about $k_{\rm B}T/2$ - the location of the transition point for the soft membrane with $n=1$. If we call $h$ the positional particle spread over the nodes (see inset of Fig.~\ref{fig4}), then the total energy cost associated with overlapping two nodes, can be estimated as 
$\bar{E}\simeq\frac{n^2}{2}\,\bar{\varepsilon}$
where  $\bar{\varepsilon}=\frac{1}{h}\int_0^h U(r)dr$. Plugging in the soft potential $U_\text{SA}$ (Eq.~\ref{usa}), and integrating we get,
\begin{equation}
\bar{\varepsilon}=\frac{\sigma\varepsilon}{2h}.  
\end{equation}
Imposing  $\bar{E}=k_{\rm B}T/2$ - the energy that sets the instability of the flat membrane -
one can estimate the number of particles per site (within the dashed circle in inset of Fig.~\ref{fig4}) when the membrane is in thermodynamic equilibrium as
\begin{equation}
n^*\approx \left[  \frac{2k_{\rm B}T}{\epsilon} \left(\frac{h}{\sigma}\right) \right]^{\frac{1}{2}}.
\end{equation}
\begin{figure}[t]
    \centering
    \includegraphics[width = 0.35\textwidth]{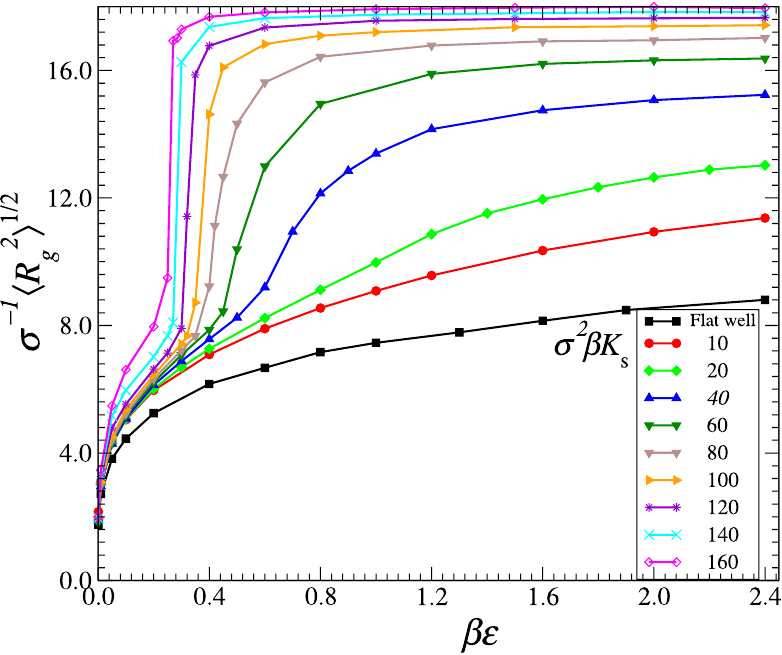}
    \caption{Radius of gyration, $R_g$, as a function of $\beta\varepsilon$ for different values of the bond stretching constant $K_s$. Here $N=30\times 30$. }
    \label{fig5}
    \vspace*{-0.5cm}
\end{figure}
This establishes the relation between the number of particles per site $n^*$ and the out of plane particle spread on a single site $h$. We now need to estimate how $h$, changes with particle loading. For this, we assume that $h$ follows the scaling for the radius of gyration of an ideal membrane, i.e.
\begin{equation}
h\sim R_g\simeq \log^{1/2}(n^*)    
\end{equation}
up to a self-avoidance of $\varepsilon\leq\varepsilon^*\simeq\frac{1}{2}k_{\rm B}T$. This functional form nicely stitches the behavior of our system for $\varepsilon=0$ (an ideal membranes/all particles are on the same site $n^*=N$) for which we expect $h=a\log^{1/2}(N)$, to the vanishing of the particle spread, $h=0$, that occurs for $\varepsilon =\varepsilon^*$, when $n^*=1$. Using this expression for $h$, we finally have 
\begin{equation}
n^*= z \log^{1/4}(n^*)    
\end{equation}
with $z\equiv(2ak_{\rm B}T/(\sigma\varepsilon))^{1/2}$. This equation can be solved  numerically to find $n^*$. 
To obtain the only remaining unknown prefactor $a$, we performed a number of simulations (not shown) of an ideal membrane ($\varepsilon=0$) as a function of its size, and fitted the data to the expected functional form $R_g=a\log^{1/2}(N)$. We find $a\simeq 3.3\,\sigma$.  Given $n^*$ blobs per site, the lateral size of the membrane scales as,  
\begin{equation}
L(n^*)\simeq (N/n^*)^{1/2}\sigma.     
\end{equation}
A plot of  $L(\varepsilon)$ vs $\varepsilon$ is shown in Fig.~\ref{theory},
and  qualitatively reproduces the behavior of $R_g$ as a function of $\varepsilon$ observed in our numerical simulations for a fixed value of $N$.
It is important to stress that as long as $(N/n^*)< h/\sigma$ (i.e. the surface length is comparable to its thickness), the surface is essentially a crumpled isotropic object, and in the limit $(N/n^*)\gg h/\sigma$, we recover the flat phase. This simple result explains the crossover behavior of $R_g$ as a function of $N$ for very small values of $\varepsilon$. In other words, no matter how small $\varepsilon$ is (except $\beta\varepsilon=0$), we can always find a sufficiently large value of $N$ for which the membrane will become flat.

\subsection{Stretching energy} 
\begin{figure}[t]
    \centering
    \includegraphics[width = 0.45\textwidth]{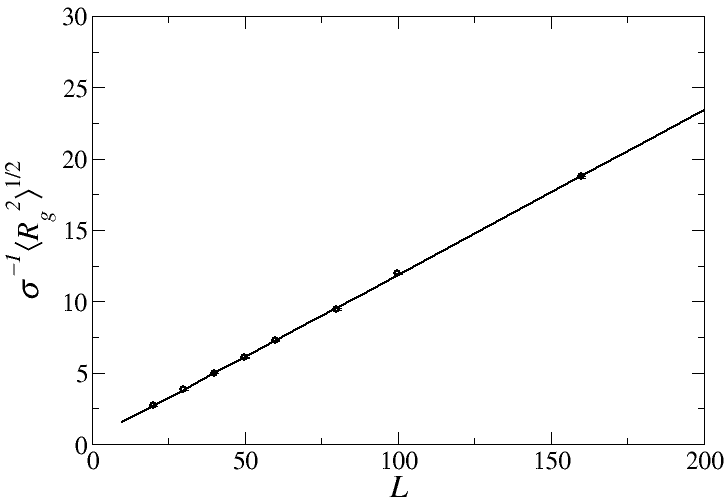}
    \caption{Radius of gyration $R_g$ as a function of the lateral size $L$ of a soft membrane. Particles are connected with entropic bonds characterized by a flat potential well $V(r)=\infty$ if $r<\sigma$ and $r>1.73\sigma$ and $V(r)=0$ otherwise. The interaction between the particles is set to $\beta\varepsilon=0.1\,k_BT$.
    The line is a fit to the data with the form $f(L)=aL^c+b$. The power $c=1.00\pm0.01$}
    \label{figD}
\end{figure}

We finally considered the effect of the stretching energy on the dependence of $R_g$ with $\varepsilon$.
Interestingly the creasing/folding drop in $R_g$ observed for $K_s=160\,k_BT\sigma^{-2}$ smoothens out as one decreases the strength of the bonding potential. As a limiting case, tagged as {\it `Flat well'} in Fig~\ref{fig5}, we also considered a model where the stretching energy was defined as a flat potential well where the bond energy is equal to zero as long as the distance between to connected particles, $\sigma\leq|r|\leq1.73\,\sigma$~\cite{kantor1986,gompper1997network}. The equilibrium structures for this model are obtained using Monte Carlo simulations \textit{via} the Metropolis algorithm, and it also shows a smooth variation of $R_g$ with $\varepsilon$. 
The vanishing of the jump in $R_g$ is due to the constraints on the surface shape a large $K_s$ imposes. Indeed for large stretching energies, the only stretch-free deformations allowed on the surface involve bending around a single axis~\cite{Witten2007Apr}, leading to folding or creasing as the only available means to reduce the area  of the surface. Systematically decreasing $K_s$ relaxes that constraint, allowing the surface to reduce its size using alternative pathways. To ensure that the specific value of $K_s$ does not affect the overall scaling of the surface with its size, we also computed the radius of gyration for the  surface with entropic {\it Flat well} bonds as a function of $N$ for $\beta\varepsilon=0.1$, and even in this case we recover a scaling that is consistent with a flat surface (see Fig~\ref{figD}). 

In Figure~\ref{figK} in the Appendix shows typical configurations of these surfaces for the largest system considered here, $N=160^2$ surface nodes, and indeed, unlike the large $K_s$ case, we observe a lack of any crystalline underlying structure in the organization of the overlapping particles. 

\section{Conclusions} 
Our numerical results  show that tethered elastic surfaces in the presence of a lattice of perforations will always be flat in the thermodynamic limit, under \textit{any} any amount of self-avoidance. This is established using an explicit polymer network model for the surface, and by exploring the limit for very large polymers  with an effective model that treats the polymers and their interactions within the blob scaling theory~\cite{Gennes1979Nov}.
Incidentally, the latter model, can also be thought of as an extension of earlier models that attempted to reduce the strength of self-avoidance by decreasing the size of the node particles in the spring and beads model~\cite{abraham1989,Boal1989Sep}. Our work extends those earlier numerical results by reaching far deeper into the ideal limit and by providing a rationale for why any degree of self-avoidance will eventually lead to the flattening of the membrane.
A crumpled phase could only be observed for degrees of self-avoidance much smaller than $k_{\rm B}T$ (which maybe  physically unrealizable), and for relatively small system sizes.
We believe these results provide crucial insight into this forty-year-old problem and establish that the shape of self-avoiding crystalline elastic surfaces as flat with or without an explicit bending rigidity or lattice perforations. 

\section*{Acknowledgements}
M.C.G acknowledges support of the Department of Atomic Energy, Government of India, under project no. RTI4001. A.C. acknowledges financial support from the National Science Foundation under Grant No. DMR-2321925. 

\bibliography{references}

\onecolumngrid
\section{Appendix}\label{appendix}
\renewcommand\thefigure{SI.\arabic{figure}} 
\setcounter{figure}{0}

\subsection{Asphericity}
The shape tensor in~\cite{rudnick_aspherity_1986} is used to quantify the morphology of our membranes,
\begin{equation}
    S_{\alpha\beta} = \frac{1}{2N^2}\sum^N_{i=1}\sum^N_{j=1}\;(r_{i\alpha}-r_{j\alpha})(r_{i\beta}-r_{j\beta})
    \label{eq:shape}
\end{equation}
$\alpha$ and $\beta$ are the $x, y$ and $z$  coordinates of a position vector $r$.
The indices $i$ and $j$ point to the identity of the particles, and the eigenvalues of the of the shape tensor $\lambda_k$ are related to the radius of gyration as $R_g^2=\sum_k\lambda_k$ and the asphericity $e$~\cite{rudnick_aspherity_1986}  is defined as $
    e \equiv \frac{3}{2}\frac{\lambda_1^2+\lambda_2^2+\lambda_3^2}{(\lambda_1+\lambda_2+\lambda_3)^2}-\frac{1}{2}$

\subsection{Overlapping Free energy of between two polymers}
The overlapping free energy between two polymers each containing $N$ monomers and radius of gyration $R_g^{(N)}$, can be estimated by considering the free energy cost associated to confining a polymer containing 2$N$ monomers  within a  spherical cavity of radius equal to that of the  radius of gyration of a single chain $R=R_g^{(N)}$. 
This can be written as $\beta \Delta F\sim (R_g^{(2N)}/R)^{\frac{3}{3\nu-1}}$~\cite{Khokhlov2002Mar,Cacciuto2006May}, where $\nu=3/5$ is the Flory scaling exponent. Using $R_{g}^{(2N)}\sim (2N)^{\nu}$ and $R\simeq R_g^{(N)}\sim N^{\nu}$, we obtain an estimate of the overlapping free energy $\beta F\sim 2^{2.25} $, which is clearly finite and independent of number of monomers. This value somewhat overestimates $2\, k_{\rm B}T$ because we neglected the fact that one polymer sharing the same space of a second polymer will likely have a slightly larger radius of gyration then if it was in isolation.

\newpage
\subsection{Morphology of the soft surface model with entropic bonds}
\begin{figure}[h]
    \centering
    \includegraphics[width = 0.47\textwidth]{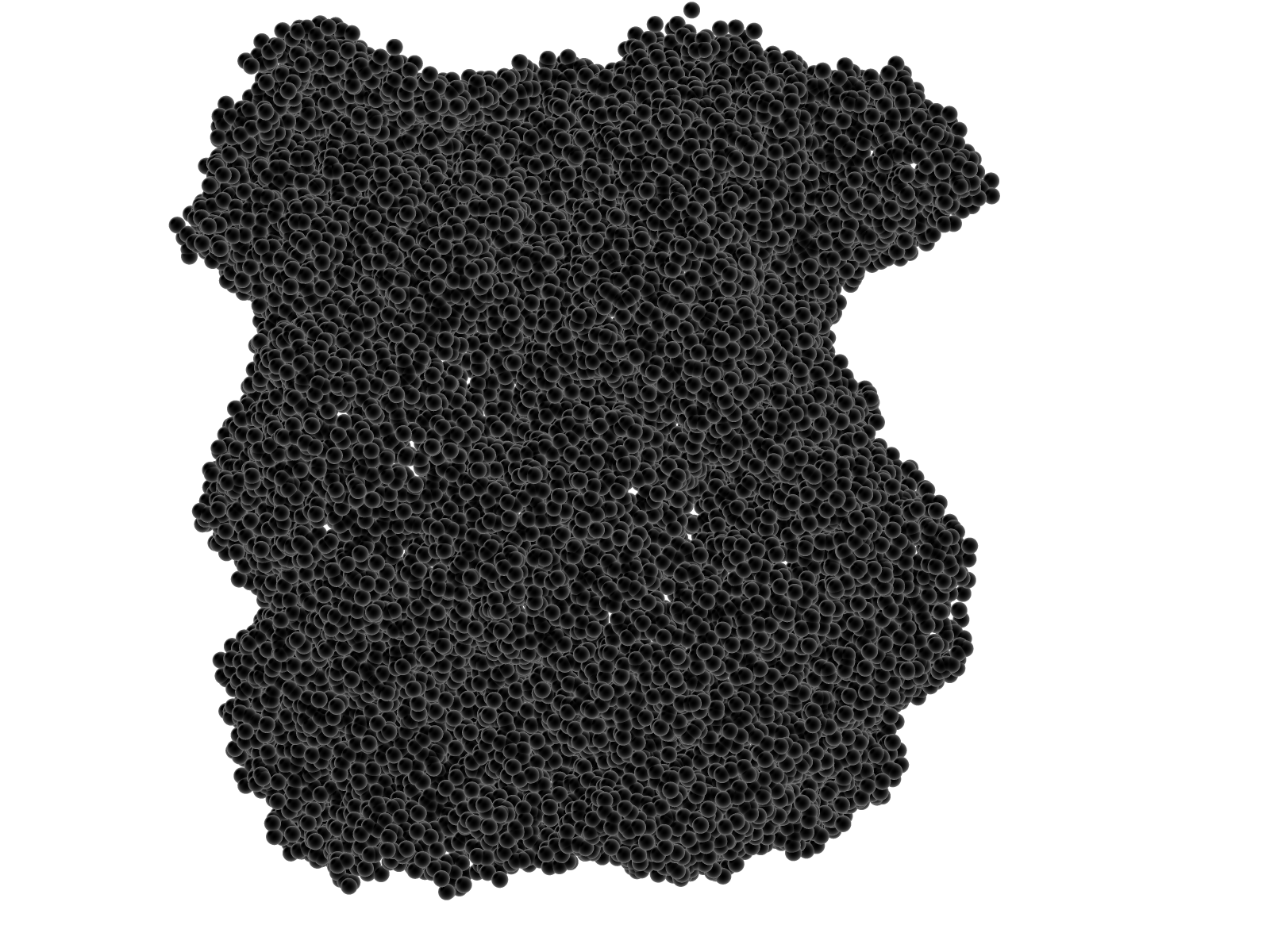}
    \includegraphics[width = 0.47\textwidth]{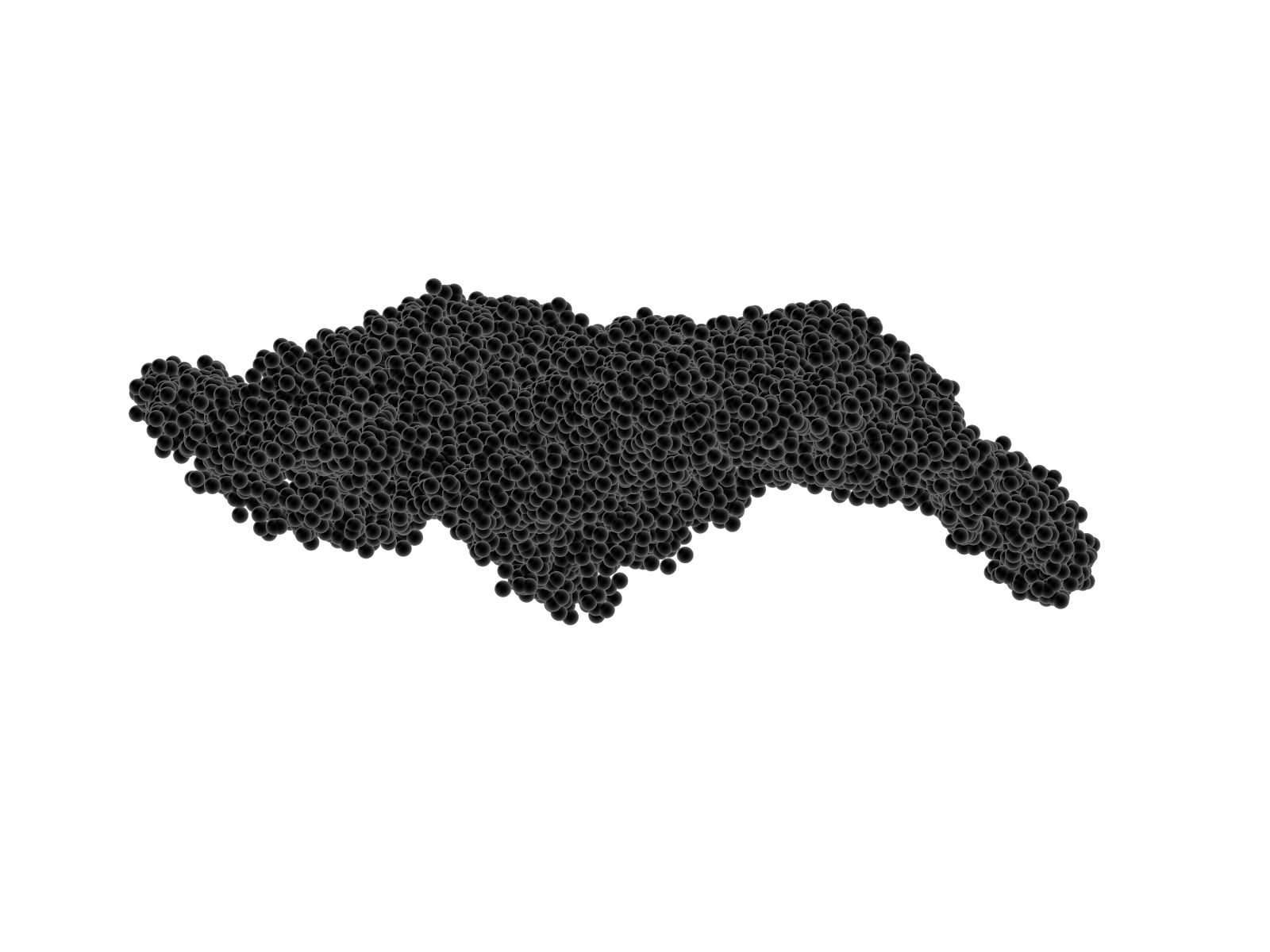}
    \includegraphics[width = 0.47\textwidth]{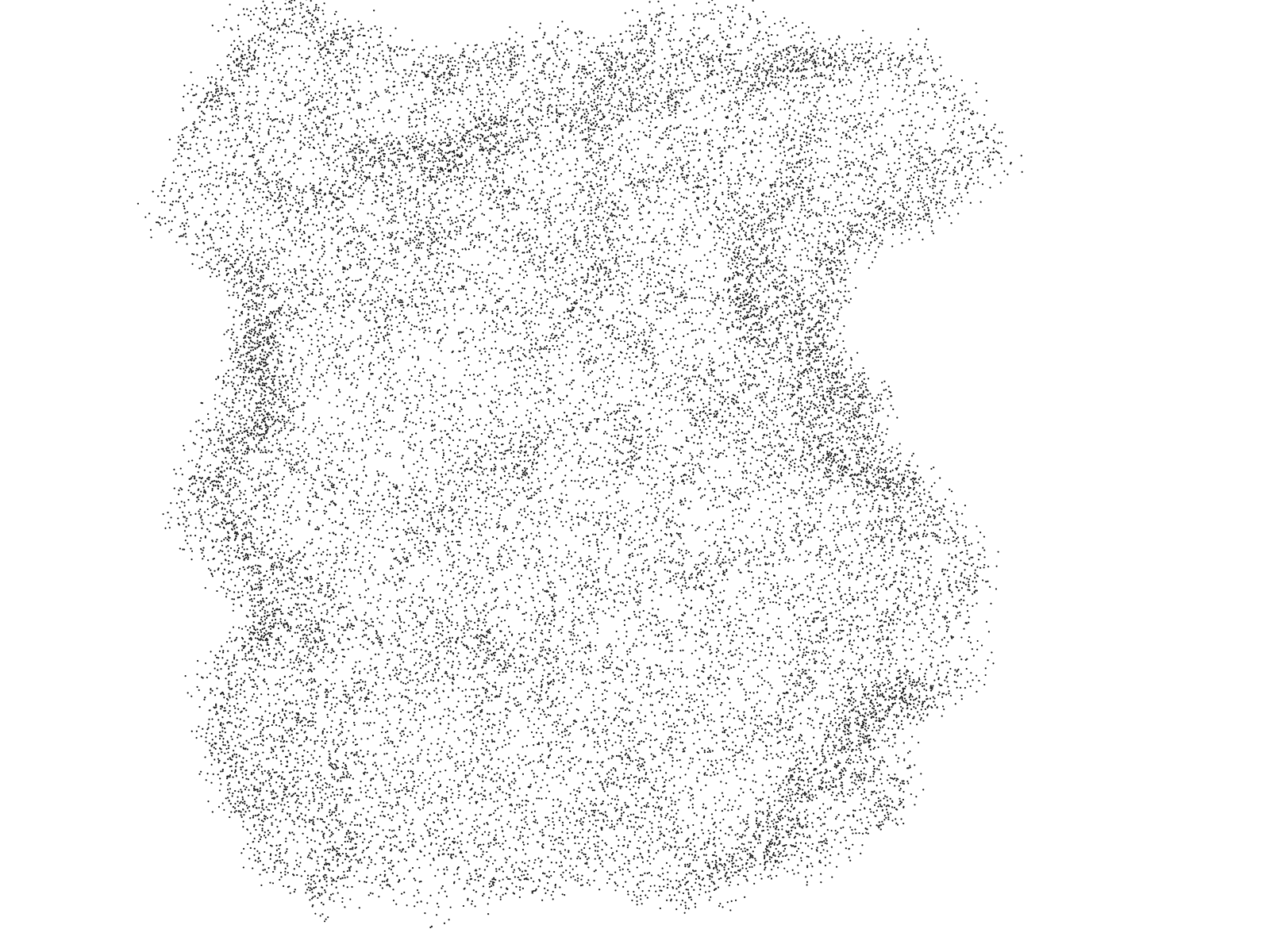}
    \includegraphics[width = 0.47\textwidth]{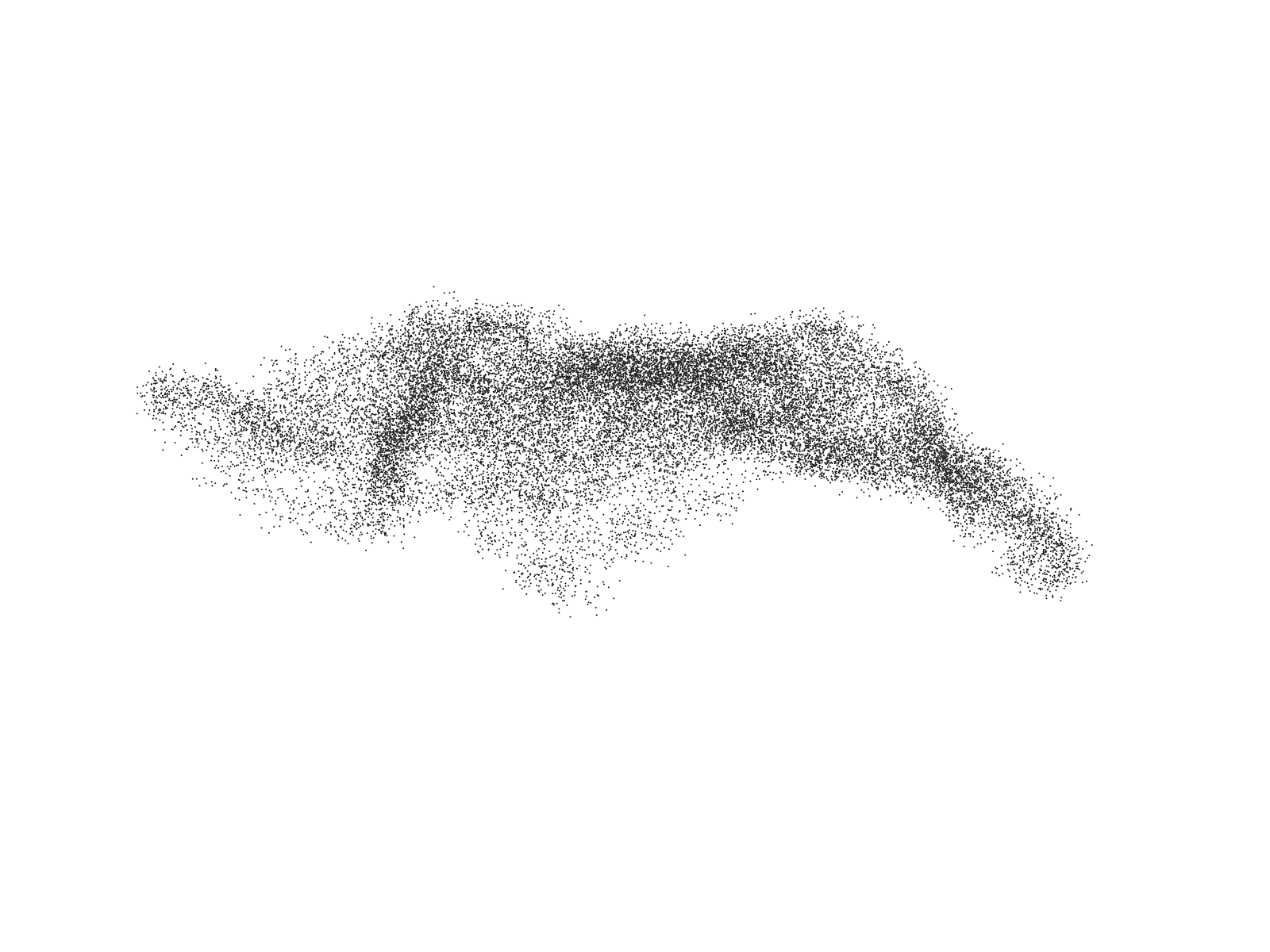}
    \caption{Snapshots from simulations of a typical configuration of a large soft membrane with $N=160^2$ particle at $\beta\varepsilon=0.1$ with stretching energies accounted for using entropic (flat well) bonds as discussed in the main text.
    The figures on top show the top view (left) and side view (right) of the surface with the actual diameter of the particles represented with black spheres.  The bottom images show the corresponding figures where the diameter has been reduced by a factor of 20 to better see deep into the membranes.
    Unlike what we found with the model with harmonic bonds for large values of the stretching energy, $K$, we see no underlaying crystalline structure in the position of the overlapping particles.}
    \label{figK}
\end{figure} 
\end{document}